\documentclass[prl,showpacs,superscriptaddress,twocolumn,longbibliography]{revtex4-1}

\usepackage{hyperref}
\hypersetup{
     colorlinks=true,
     linkcolor=magenta,
     filecolor=blue,
     citecolor=blue,      
     urlcolor=cyan,
     }

\usepackage{hyperref}
\usepackage{color}
\usepackage[usenames,dvipsnames]{xcolor}
\usepackage{amsmath,amsthm,amssymb}
\usepackage{graphicx}
\usepackage{float}
\usepackage{epsfig}
\usepackage{bm}% bold math
\usepackage{mathrsfs}
\usepackage{multirow}
\usepackage[all]{xy}
\usepackage{pbox}
\usepackage{verbatim}
\usepackage{braket}
\usepackage{mathtools}
\usepackage{bm}
\usepackage{tikz}
\usepackage{xcolor}
 
\usepackage{mathtools}

\usepackage{mathtools}
\usepackage{enumerate}   

\DeclareMathOperator{\Tr}{Tr}

\newcommand{\er}[1]{Eq.~\eqref{#1}}
\newcommand{\ers}[2]{Eqs.~(\ref{#1}-\ref{#2})}

\newcommand{\beq}{\begin{equation}}
\newcommand{\eeq}{\end{equation}}

\begin{document}  

\title{Finite time large deviations via matrix product states}

\author{Luke Causer}
\affiliation{School of Physics and Astronomy, University of Nottingham, Nottingham, NG7 2RD, UK}
\affiliation{Centre for the Mathematics and Theoretical Physics of Quantum Non-Equilibrium Systems,
University of Nottingham, Nottingham, NG7 2RD, UK}
\author{Mari Carmen Ba\~nuls}
\affiliation{Max-Planck-Institut f\"ur Quantenoptik, Hans-Kopfermann-Str.\ 1, D-85748 Garching, Germany}
\affiliation{Munich Center for Quantum Science and Technology (MCQST), Schellingstr.\ 4, D-80799 M\"unchen}
\author{Juan P. Garrahan}
\affiliation{School of Physics and Astronomy, University of Nottingham, Nottingham, NG7 2RD, UK}
\affiliation{Centre for the Mathematics and Theoretical Physics of Quantum Non-Equilibrium Systems,
University of Nottingham, Nottingham, NG7 2RD, UK}

\begin{abstract}
Recent work has shown the effectiveness of tensor network methods for computing large deviation functions in constrained stochastic models in the infinite time limit. 
Here we show that these methods can also be used to study the statistics of dynamical observables at {\em arbitrary finite time}. This is a harder problem because, in contrast to the infinite time case where only the extremal eigenstate of a tilted Markov generator is relevant, for finite time the whole spectrum plays a role. We show that finite time dynamical partition sums can be computed efficiently and accurately in one dimension using matrix product states, and describe how to use such results to generate rare event trajectories on demand. We apply our methods to the Fredrickson-Andersen (FA) and East kinetically constrained models, and to the symmetric simple exclusion process (SSEP), unveiling dynamical phase diagrams in terms of counting field and trajectory time. We also discuss extensions of this method to higher dimensions. 
\end{abstract}

\maketitle

\noindent {\bf\em Introduction.--}
Large deviation (LD) theory provides a powerful framework to investigate the statistical fluctuations of time-averaged observables in stochastic systems (for reviews, see e.g. Refs.~\cite{Touchette2009, Garrahan2018, Jack2020, Limmer2021}).
At long times (assuming finite correlation times)   
the probabilities of such observables obey a LD principle, and the corresponding scaled cumulant generating function (SCGF, see below) can be retrieved from the leading eigenvalue of the \emph{tilted} (or deformed or biased) generator \cite{Touchette2009}.
For large systems, estimating this eigenvalue is difficult, so one resorts to sampling the corresponding biased trajectory ensemble via numerical methods such as trajectory importance sampling \cite{Bolhuis2002, Ray2018, Klymko2018, Guyader2020}, population dynamics \cite{Borkar2003, Giardina2006, Lecomte2007b}, optimal control \cite{Jack2010, Nemoto2016, Ferre2018, Ray2018b, Jacobson2019, Das2019, Ray2020}, or machine learning approaches \cite{Oakes2020, Rose2021, Casert2020, Whitelam2020, Das2021, Yan2021}. For lattice models, recent work has focused on the use of tensor network (TN) techniques to approximate the leading eigenvector of the tilted generator through variational means \cite{Banuls2019, Helms2019, Causer2020} or power methods \cite{Helms2020}.

A harder problem is that of computing the statistics of time-averaged observables for {\em finite time} \cite{Nemoto2017b, Hidalgo2017, Baek2019}. The reason is that away from the long time limit the corresponding dynamical partition sums (i.e., moment generating functions) do not obey a LD principle in time - only obeying an LD principle in space for large sizes - and as a consequence they are not determined only by the leading eigenvalue of the tilted generator, but by their whole spectrum. If time is very short, one can get away with direct sampling, but for intermediate times the usual sampling approaches fall short \cite{Causer2021}.
Here we develop a scheme to study these rare events by implementing well-developed TN techniques to simulate time evolution. This allows us to calculate dynamical partition functions for trajectories of arbitrary time extent. Furthermore, we show how to use the results here to directly simulate stochastic trajectories in finite-time tilted ensembles at small computational cost, thus generalising the method of Ref.~\cite{Causer2021}.

We focus for concreteness on one dimensional kinetically constrained models (KCMs) - often used in the modelling of structural glasses \cite{Ritort2003, Chandler2010, Garrahan2011, Garrahan2018} - specifically the Fredrickson-Andersen (FA) \cite{Fredrickson1984} and the East \cite{Jackle1991} models, and on the symmetric simple exclusion process (SSEP). Both KCMs and SEPs display phase transitions in their dynamical LDs in the long-time limit \cite{Garrahan2007, Bodineau2007, Appert-Rolland2008, Garrahan2009, Bodineau2012, Jack2015, Karevski2017}. With the methods developed here we are able to construct the dynamical phase diagram both as a function of counting field and of trajectory time, determining finite time scaling of active-inactive phase transitions in these models, and uncovering the emergence with time of the correlated structure of the active phase in the East model and the SSEP.

\smallskip

\noindent {\bf\em Models.--}
The three models we consider live in a one dimensional lattice of $N$ sites, with binary variables $n_{j}=0, 1$ for each $j = 1\cdots N$, evolving under continuous-time Markov dynamics with local transitions. The probability for each configuration $\ket{x} = \ket{n_{1}\cdots n_{N}}$ at time $t$, encoded in a vector $\ket{P(t)} = \sum_{x}P(t, x)\ket{x}$, evolves deterministically via a Master equation, 
$\partial_t \ket{P(t)} = \mathbb{W} \ket{P(t)}$, where $\mathbb{W}$ is the Markov generator. Being a stochastic operator $\mathbb{W}$ has a structure $\mathbb{W} = \mathbb{K} - \mathbb{R}$, with an off-diagonal matrix of transition rates $\mathbb{K}$, and a positive diagonal matrix of escape rates $\mathbb{R}$. 

For the KCMs the generator reads
\begin{align}
    \mathbb{W}^{\rm KCM} = \sum_{i} f_{i} \big[ &c\sigma_{i}^{+} + (1-c)\sigma_{i}^{-} 
    \nonumber \\
    &- c (1-n_{i}) - (1-c)n_{i} \big],
    \label{W}
\end{align}
where $c\in(0, 1/2]$ defines the site occupation at equilibrium, and $\sigma_{i}^{\pm}$ are the Pauli raising and lowering operators at site $i$. 
Spin flips are only permitted if the kinetic constraint, $f_{i}$, is satisfied.
We consider two paradigmatic KCMs, the Fredrickson-Andersen (FA) \cite{Fredrickson1984} model and the East \cite{Jackle1991} model. They are defined by the respective constraint functions
\begin{align}
    f_{i}^{{\rm FA}} &= n_{i-1} + n_{i + 1} , 
    \;\; 
    f_{i}^{{\rm East}} = n_{i-1} .
\end{align}
We consider lattices with open boundary conditions (OBC) to allow for efficient tensor network contractions. For numerical convenience, we choose the fixed boundaries $n_{1} = n_{N} = 1$ for the FA {\footnote{This choice of boundaries has no spontaneous breaking of symmetry at $s\gg0$.}} model and $n_{1} = 1$ for the East model.
The corresponding stationary states are product states, 
\begin{align}
    \ket{{\rm ss^{FA}}} &= \ket{1} \otimes \left[(1-c)\ket{0} + c\ket{1}\right]^{\otimes N-2} \otimes \ket{1},
    \\
    \ket{{\rm ss^{East}}} &= \ket{1}\otimes\left[(1-c)\ket{0} + c\ket{1}\right]^{\otimes N-1}.
\end{align}

The third model we consider is the symmetric simple exclusion process (SSEP) whose generator reads
\begin{align}
    \mathbb{W}^{\rm SSEP} = \frac{1}{2} \sum_{i} \big[ &
    	\sigma_{i}^{+} \sigma_{i+1}^{-} + 
    	\sigma_{i}^{-} \sigma_{i+1}^{+} 
	\label{SSEP}
	\\ & \;\;\;\;	
		- 
		(n_{i}+n_{i+1}) + 2 n_{i} n_{i+1} \big]
		\nonumber
\end{align}
For the SSEP we consider OBC such that particles can enter and leave at the boundaries with rate $1/4$. The stationary state is $\ket{\rm ss^{\rm SSEP}} = 2^{-N}\ket{-} = 2^{-N}\sum_{x} \ket{x}$, with the ``flat'' state $\bra{-}$ being the leading left eigenvector of each generator above.

\smallskip

\noindent {\bf\em Dynamical rare events and LDs.--}
We now consider the ensemble of all possible trajectories $\{\omega_{\alpha}\}$ with trajectory time $t$, where $\omega_{\alpha} = \{x_{0} \to x_{t_{1}} \to \cdots \to x_{t}\}$ defines jumps to configurations $x_{t_{k}}$ at times ${t_{k}}$. 
The probability of observing the value $K(\omega_{\alpha}) = K$ of some time-integrated observable $K$ is 
\beq
P_{t}(K) = \sum_{\alpha} \pi(\omega_{\alpha})\delta(K(\omega_{\alpha}) - K) ,
\label{P}
\eeq
where $\pi(\omega_\alpha)$ defines the probability of observing the trajectory. The corresponding moment generating function (or trajectory partition sum) is
\beq
    Z_{t}(s) = \sum_{\alpha} \pi(\omega_{\alpha})e^{-sK(\omega_{\alpha})}
    \label{Z}
\eeq
where the \emph{counting field} $s$ is conjugate to the observable. 

For large times, both \ers{P}{Z} take a large deviation (LD) form in time \cite{Touchette2009,  Garrahan2009, Garrahan2007, Lecomte2007}, $P_{t}(K) \asymp e^{-t\varphi(K/t)}$ and $Z_{t}(s) \asymp e^{t\theta(s)}$. The LD rate function $\varphi(K/t)$ and the scaled cumulant generating function (SCGF) $\theta(s)$ play the roles of a trajectory entropy density and a free-energy density, respectively, and are related through the Legendre transform $\theta(s) = -\min_{k}\left[sk+ \varphi(k)\right]$. 

The partition sum \er{Z} can be written as
\beq
    Z_{t}(s) = \braket{- | e^{t\mathbb{W}_{s}} | {\rm ss}}
    \label{ZW}
\eeq
in terms of the {\em tilted generator} $\mathbb{W}_{s}$
\cite{Touchette2009, Garrahan2007, Lecomte2007, Garrahan2009}.
In what follows we focus on the {\em dynamical activity} \cite{Garrahan2007, Lecomte2007, Garrahan2009, Maes2020}), that is, the total number of spin flips, as a trajectory observable. In this case, $\mathbb{W}_{s} = e^{-s}\mathbb{K} - \mathbb{R}$. While for large times all that is needed to determine \er{ZW} is the dominant eigenstate of $\mathbb{W}_{s}$, for  
finite times the whole spectrum of $\mathbb{W}_{s}$ is required.

\begin{figure}[t]
    \centering
    \includegraphics[width=\linewidth]{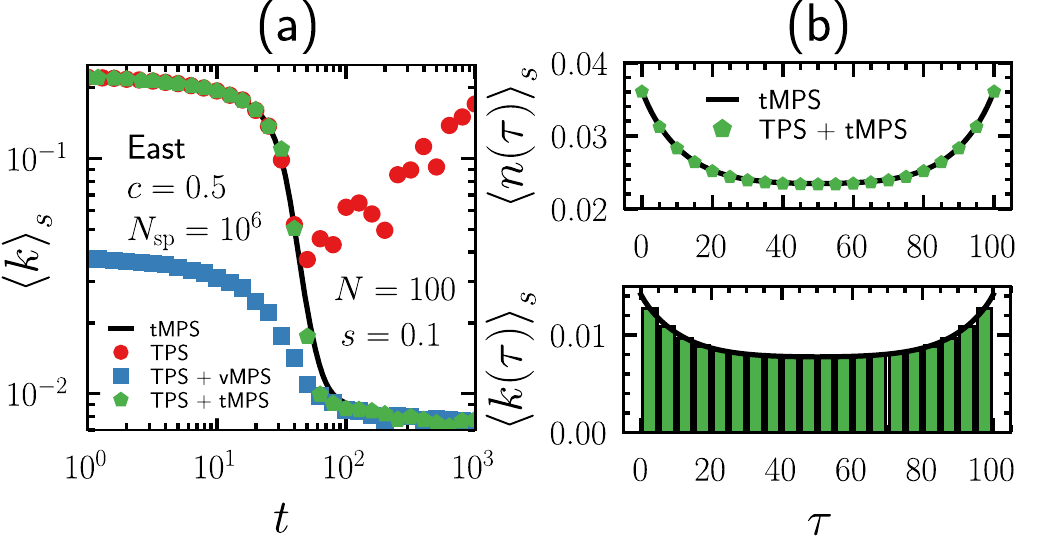}
    \caption{\textbf{Demonstration of the methods.} 
    East model at $c = 0.5$, $N = 100$ and $s = 0.1$.
    (a) Dynamical activity $\braket{k}$ from {tMPS} (black line), TPS with no auxiliary dynamics (red circles), TPS with the LD eigenvector auxiliary dynamics {via vMPS} (blue squares), and TPS with a {tMPS} reference dynamics (green pentagons).
    (b) Time-dependent occupations (top) and instantaneous activity (bottom) from MPS time-evolution (black line) from direct sampling with a {tMPS} auxiliary dynamics (green pentagons / bars).
    }
    \label{fig: method}
\end{figure}

\begin{figure*}[t]
    \centering
    \includegraphics[width=1\linewidth]{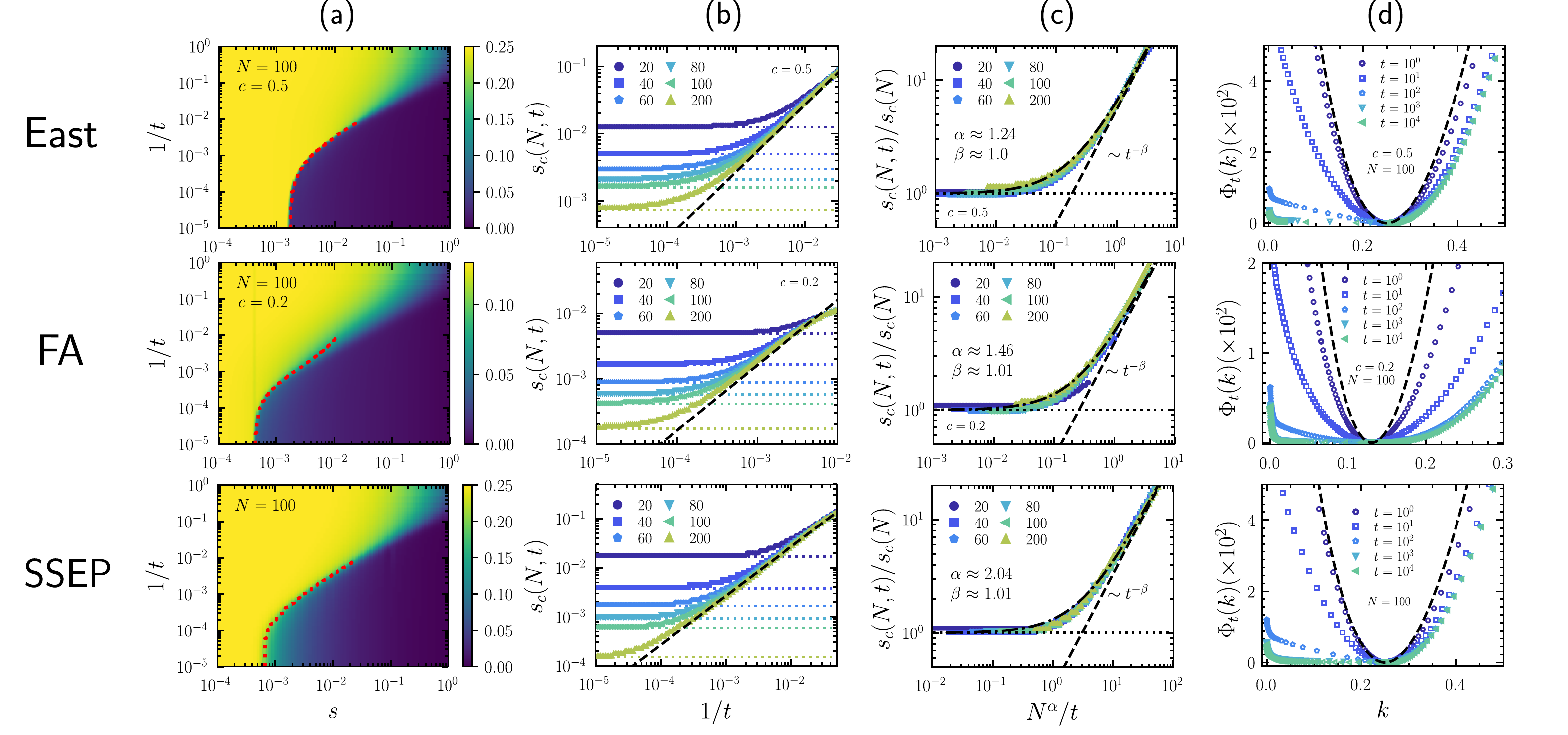}
    \caption{\textbf{Rare event statistics.} 
    The rare event statistics for the East model with $c = 0.5$ (top) and the FA model with $c = 0.2$ (middle) and SSEP (bottom).
    (a) The dynamical activity $k(s, t)$ as a function of $s$ and inverse time $1/t$ for $N = 100$. The red dotted line indicates our estimate of the transition point.
    (b) The transition point for various system sizes $N\in[10, 200]$. The dotted lines indicate the infinite time value (see Refs.~\cite{Banuls2019, Causer2020}), and the dashed line shows $s_{c}(N, t) \sim t^{-1}$.
    (c) The same data is shown in (b) but with $s_{c}(N, t)$ scaled by $s_{c}(N)$ and time scaled by $N^{-\alpha}$, where $\alpha$ is the critical exponent extracted from $s_{c}(N)$. The dotted line shows where the y-axis is one, and the dashed line shows $t^{-\beta}$. The sum of both lines is given by the dashed-dotted line.
    (d) The estimate of the rate function $\Phi_{t}(k)$ defined in \er{rate}. The dashed line shows a Poisson distribution with the equilibrium average as its mean.
    All of the data is calculated using the dynamical partition sum $Z_{t}(s)$ from tMPS.
    }
    \label{fig: scaling}
\end{figure*}

\smallskip
\noindent {\bf\em Finite time statistics from MPS.--}
The models we consider obey detailed balance. This allows us to write $\mathbb{W}_{s}$ in a Hermitian form through a similarity transformation independent of $s$ \cite{Garrahan2009}, $\mathbb{H}_{s} = \mathbb{P}^{-1/2}\,\mathbb{W}_{s}\,\mathbb{P}^{1/2}$, where $\mathbb{P}^{1/2}$ is a diagonal matrix of probability amplitudes at equilibrium (for the SSEP, $\mathbb{P}$ is the identity). As a consequence the leading eigenvalue of $\mathbb{H}$ obeys a Rayleigh-Ritz variational principle, {allowing the application of variational methods} such as the density matrix renormalisation group (DMRG) \cite{White1992}. We then write \er{ZW} as
\beq
    Z_{t}(s) = \braket{\psi_{0} | e^{t\mathbb{H}_{s}} | {\rm \psi_{0}}},
    \label{ZH}
\eeq
where $\ket{\psi_{0}} = \mathbb{P}^{-1/2}\ket{{\rm ss}} = \left[\bra{-} \mathbb{P}^{1/2}\right]^{\dag}$.
It is useful to define the time evolved vector $\ket{\psi_{\tau}} = e^{\tau\mathbb{H}_{s}}\ket{\psi_{0}}$ ($\tau \leq t$).
The partition function can then be written as $Z_{t}(s) = \braket{\psi_{t-\tau} | \psi_{\tau}}$, and in particular can be determined by only evolving the vector by $\tau = t / 2$.

The average dynamical activity (per unit time and site) of the biased ensemble of trajectories follows from the partition sum,
\beq
    k(s) = -\frac{1}{Nt}\frac{d}{ds} \log(Z_{t}(s)).
    \label{k}
\eeq
We can also calculate time-dependent configurational observables for any $0 \leq \tau \leq t$,
\begin{align}
    \braket{O(\tau)}_{s} &= Z_{t}(s)^{-1} \braket{\psi_{0} | e^{(t-\tau)\mathbb{H}_{s}} O e^{\tau\mathbb{H}_{s}} | \psi_{0}}
    \nonumber
    \\
    &= Z_{t}(s)^{-1} \braket{ \psi_{t-\tau}| O | \psi_{\tau}}.
    \label{O}
\end{align}

In order to compute the time-evolved state $\ket{\psi_{t}}$ we use methods from quantum many-body {physics}, in particular, matrix product states (MPS) (for reviews, see Refs.~\cite{Schollwoeck2011, Orus2019}) \footnote{Tensor network calculations were implemented using the ITensor library \cite{itensor}}. Here we use both {variational optimization of MPS (vMPS, e.g. \cite{Schollwoeck2011, Banuls2019}), and time-evolved MPS (tMPS, e.g. \cite{Verstraete2004b}).
Notice that for long times, $\ket{\psi_{\tau}}$ becomes close to the leading eigenvector of $\mathbb{W}_{s}$. We exploit this fact to simulate evolution for long times with higher precision, see the Supplemental Material (SM) \footnote{See Supplemental Material for a description of the numerical methods, which includes \cite{Suzuki1976, Paeckel2019}.} for details.}

In Ref.~\cite{Causer2021}, we used the MPS approximation (from vMPS) to the ground state of $\mathbb{H}_{s}$ to construct a near-optimal dynamics, which when supplemented with trajectory importance sampling (specifically transition path sampling or TPS \cite{Bolhuis2002}), allowed us to efficiently simulate trajectories in large time tilted ensembles.
Here we apply {the same scheme, but} instead use the time-evolved state $\ket{\psi_{t/2}}$. We construct a time-independent dynamics which approximates the optimal (or Doob) dynamics at the centre of finite-time trajectories under tilting \footnote{It is possible also to approximate dynamics at any point in the trajectory. For large times, the state evolved to time $\tau = t/2$ is however the most efficient.}.

Figure \ref{fig: method}(a) compares various sampling {methods} in the East model at $s>0$. The dynamics is active at short times (due to initial conditions) and inactive at large times \cite{Garrahan2007, Garrahan2009, Banuls2019}. We show the activity from the partition sum calculated via MPS time-evolution (black line) as a function of trajectory length. We also show sampling with TPS with the original dynamics (red circles); this method only accounts for the dynamical activity $\braket{k}$ at short times, and fails at long times. 
The methods introduced in Ref.~\cite{Causer2021} construct the long-time optimal (Doob) dynamics with the approximate leading eigenstate from vMPS. We then apply TPS with this dynamics to sample trajectories for arbitary time. This accounts for $\braket{k}$ at long times \cite{Causer2021}, but fails at short times.
If we adopt this method, but replace the MPS used in the auxiliary dynamics with the time evolved state (green pentagons) we get accurate results for the activity for all trajectory lengths. Despite the fact that the exact Doob dynamics for finite time is in general time-dependent \cite{Chetrite2015}, this latter approach with a time-independent dynamics for each trajectory length $t$ is efficient enough for  TPS to converge to the actual finite-time tilted ensemble, thus correcting any discrepancies. In the SM we provide a detailed comparison. 
Figure~\ref{fig: method}(b) shows the averaged time-dependant occupations $\braket{n(\tau)}_{s}$ (top) and instantaneous activity $\braket{k(\tau)}_{s}$ (bottom) for some fixed trajectory time $t=100$, generated from the $s$-ensemble at $s = 0.1$ for both tMPS and tMPS+TPS.

\begin{figure*}[t]
    \centering
    \includegraphics[width=\linewidth]{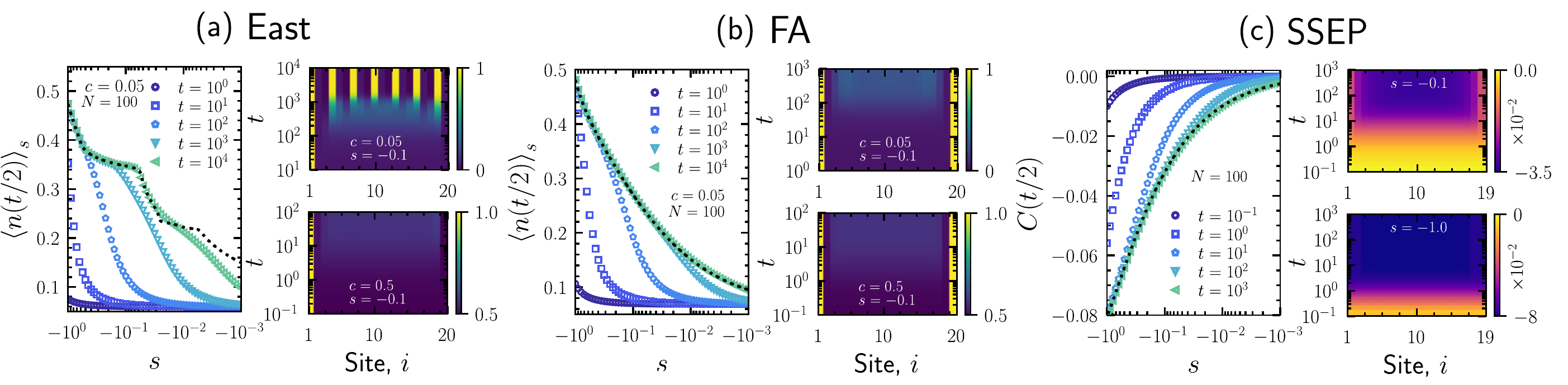}
    \caption{\textbf{Structures in the active phase.} 
    We show the average occupations at the center of the trajectory $\braket{n_{i}(t/2)}_{s}$ for $s = -0.1$ for the (a) East model and (b) FA models.
    The left panels of each show the lattice average for a range of $s$ and $t$ with $c = 0.05$, whilst the right panels show the occupations at each site for $c = 0.05$ (top) and $c = 0.5$ (bottom), with $s = -0.1$.
    We show the same for the SSEP in (c) but with the nearest neighbour correlations $C_{i}(t/2)$. The right panels are for $s=-0.1$ (top) and $s=-1.0$ (bottom).
    Dotted lines show the expected value at infinite times.
    All observables are calculated from the time-evolved MPS $\ket{\psi_{t/2}}$.}
    \label{fig: structures}
\end{figure*}

\smallskip
\noindent {\bf\em Finite time scaling of active-inactive transition.--}
The three models we study here display an LD phase transition  \cite{Garrahan2007, Appert-Rolland2008, Garrahan2009, Jack2015} in the long time and large size limit between a dynamical phase where activity is extensive in space and one where activity is subextensive. The finite size scaling analysis of this transition in the long time limit has been studied theoretically \cite{Appert-Rolland2008, Bodineau2012, Bodineau2012b, Nemoto2017, Jack2020b} and numerically \cite{Nemoto2017, Banuls2019}: for finite size the active-inactive transition is smoothed into a sharp crossover located at $s_{c}(N, t=\infty)>0$, which decreases as an inverse power of the system size. In general, however, the location of the transition point depends both on time and size, $s_{c}(N, t)$, but a detailed numerical analysis of the finite time scaling has not been possible to date due to the difficulty of simulating efficiently rare trajectories at intermediate times \cite{Causer2021}. With the approach presented above we can now investigate this issue in detail.

Figure \ref{fig: scaling}(a) shows the 
dynamical activity $k(s)$ 
as a function of $s$ and inverse time $t^{-1}$ (East model, top row; FA model, middle; SSEP, bottom). There is a transition from a high activity (light) to low activity (dark) as $s$ is increased which becomes sharper and moves to smaller $s$ with increasing time. The point $s_{c}(N, t)$ (shown by the red dashed line) is that of the peak in the dynamical susceptibility $\chi(s, t) = dk(s)/ds$. These dynamical phase diagrams are reminiscent of those of (first-order) quantum phase transitions \cite{Vojta2003}, with $s$ as an applied field and the inverse time as temperature. 

The scaling of the transition point is shown as a function of (inverse) time for multiple system sizes $N\in[20, 200]$ in column (b) of Fig.~\ref{fig: scaling}. For small times the transition point scales approximately as $s_{c} \sim t^{-1}$ for the three models. When time becomes large enough finite-size effects start to dominate.
{For simplicity, we use the approximate form}
\begin{align}
    s_{c}(N, t) \approx s_{c}(N) + s_{c}(t) ,
\end{align}
where $s_{c}(N) \sim N^{-\alpha}$ can be extracted from {vMPS \cite{Banuls2019, Causer2020}}. For the FA and East models the exponent $\alpha > 1$ \cite{Banuls2019}, while for the SSEP we find the expected $\alpha \approx 2$ \cite{Appert-Rolland2008}. 
In column (c) of Fig.~\ref{fig: scaling} we show how the $s_{c}(N, t)$ curves can be collapsed, allowing us to estimate $s_{c}(t) \sim t^{-\beta}$. We find $\beta \approx 1$ for all models.

Also important to the rare event statistics is the probability distribution of the dynamical activity, $P_{t}(K)$. While for finite times
$Z_{t}(s)$ and $P_{t}(K)$ do not obey a LD principle in time, for large sizes they still obey one in system size, $Z_{t}(s) \asymp e^{N \Theta(s,t)}$ and $P_{t}(K) \asymp e^{- N \Phi_t(K/N)}$. We can therefore obtain the time-dependent rate function $\Phi_t(K)$ through the Legendre transform
\begin{align}
    \Phi_t(k) &= - \max_s
    \left[\Theta(s, t) + s k \right] , 
    \label{rate}
\end{align}
where $\Theta(s,t) = N^{-1} \log Z_{t}(s)$. From the numerical estimate of $Z_{t}(s)$ we can therefore estimate $\Phi_t(K/N)$ for all times. Column (d) of  Fig.~\ref{fig: scaling} shows the corresponding rate functions for system size $N = 100$. For small times, the distribution of the activity is close to Poissonian (dashed line), in agreement with the absence of a transition. As time increases the rate function broadens into the 
characteristic shape of a first order phase transition \cite{Garrahan2007, Banuls2019}.

\smallskip
\noindent {\bf\em Structure of the active phase.--}
Long time trajectories with an atypically large activity are known to display an interesting structure in two of the models we consider here \cite{Jack2013, Jack2015, Banuls2019}. Our finite time method allows to study how such structure depends on the trajectory length. 

{
In Fig.~\ref{fig: structures}(a), we show the average occupation at the mid point of the dynamics. The left panel shows the lattice averaged occupations $\braket{n(t/2)}_{s}$ at time $\tau = t/2$ for ensembles of trajectories with total time $t$, as a function of $s$ for various $t$, at $c=0.05$. The panels on the right show the average spatial profile $\braket{n_{i}(t/2)}_{s}$ at $s=-0.1$ for 
$c = 0.05$ (top) and $c = 0.5$ (bottom). In both cases, the average density is spatially featureless at short times, but arranges to maximise activity at long times. For $c = 0.05$ it does so by forming an anti-correlated structure, while these anti-correlations are absent for $c = 0.5$ (cf.\ the long time case \cite{Jack2013}). Figure~\ref{fig: structures}(b) shows the same for the FA model, where there is no appreciable structure forming for small $c$. Notice also from the left panels the longer times needed to reach the LD behaviour in the East compared to the FA model.

In Fig.~\ref{fig: structures}(c) we quantify the local structure of the SSEP in terms of the nearest-neighbour correlations
\beq
    C_{i}(t) = \braket{n_{i}n_{i+1}(t)}_{s} - \braket{n_{i}(t)}_{s} \braket{n_{i+1}(t)}_{s},
\eeq
and the lattice average $C(t) = (N-1)^{-1}\sum_{i=1}^{N-1} C_{i}(t)$.
The right panels show a growth of anti-correlated order with increasing trajectory length towards the ``hyperuniform'' arrangement at long times, cf.\ Ref.~\cite{Jack2015}.

\smallskip
\noindent {\bf\em Conclusions.--}
We have implemented a time evolution scheme using MPS to study the rare events of one dimensional KCMs in finite-time trajectories. In this way we have extended recent efforts on the long-time LD statistics via TNs to the arguably harder problem of the LDs away from the long time limit. We showed how to directly compute dynamical partition sums, and we derived an efficient sampling scheme for finite-time rare trajectories.
Understanding the finite-time behaviour of dynamical systems is significant, as the times required to observe long-time behaviour can be too large to implement experimentally.
A next step would be to extend these ideas to dimensions larger than one. A possibility could be to implement sampling through two dimensional TNs, such as PEPS (e.g. \cite{Verstraete2004c, Verstraete2008}), which have already proven useful in studying the LDs in the long-time limit of two dimensional exclusion processes \cite{Helms2020}. While bond dimensions will be limited in this case, using a time evolution scheme like we presented here one could approximate the reference dynamics for the centre of trajectories (i.e. evolve by $e^{t\mathbb{W}_{s}/2}$) alongside a scheme such as TPS to obtain reliable results.
Another direction would be to apply the methods demonstrated here to driven problems, such as currents in exclusion processes. Here we cannot exploit hermiticity, and would have to compute the time-evolved left and right eigenvectors.
We hope to report on such studies in the near future.

\bigskip

\noindent {\bf\em Acknowledgements.--}
We acknowledge financial support from EPSRC Grant no.\ EP/R04421X/1 and the Leverhulme Trust Grant No. RPG-2018-181. 
M.C.B.\ acknowledges support from Deutsche Forschungsgemeinschaft (DFG, German Research Foundation) under Germany's Excellence Strategy -- EXC-2111 -- 390814868.
We acknowledge access to the University of Nottingham Augusta HPC service.

\bibliographystyle{apsrev4-1}
%\bibliography{biblography}

\newpage
\onecolumngrid
\section{Supplemental Material} 
% To be included in main file or otherwise

\section{Methods}

The vector $\ket{\psi_{t}}$ is given as a  matrix product state (MPS) ansatz,
\begin{equation}
    \ket{\Psi} = \sum_{i_{1}\cdots i_{N}}^{d} \Tr(A_{1}^{i_{1}}A_{2}^{i_{2}}\cdots A_{N}^{i_{N}}) \ket{i_{1}\, i_{2} \, \cdots \, i_{N}}
    \tag{S1}
\end{equation}
where each $A_{j}^{k}$ is a rank-3 tensor with dimensions $(d, D, D)$, with the variational parameter $D$ (known as the \emph{bond dimension}) and $k=1\dots d$.
We implement time evolution using a hybrid approach. 
We start with the equilibrium steady state, $\ket{\psi_{0}}$, which can be written as a product state (an MPS with $D = 1$),
\begin{align}
    \ket{\psi_{0}^{\rm FA}} &= \ket{1} \otimes \left[ \sqrt{1-c}\ket{0} + \sqrt{c}\ket{1} \right]^{\otimes N-2} \otimes \ket{1},
    \tag{S2}
    \\
    \ket{\psi_{0}^{\rm East}} &= \ket{1} \otimes \left[ \sqrt{1-c}\ket{0} + \sqrt{c}\ket{1} \right]^{\otimes N-1}.
    \tag{S3}
\end{align}
The first step is to find the leading eigenvector $\ket{\psi^{\rm LD}}$ of $\mathbb{H}_{s}$.
This can be achieved by employing variational MPS (vMPS, see e.g. Ref.~\cite{Banuls2019} for details).
We then project the initial state $\ket{\psi_{0}}$ onto the (unnormalized) LD vector and its orthogonal complement,
\begin{align}
    \ket{\psi_{0}^{\rm LD}} &= \mathcal{P} \ket{\psi_{0}} ,
    \tag{S4}
    \\
    \ket{\psi_{0}^{\rm rem}} &= \left(1-\mathcal{P}\right) \ket{\psi_{0}} ,
    \tag{S5}
\end{align}
where $\mathcal{P} = \ket{\psi^{\rm LD}}\bra{\psi^{\rm LD}}$. 
The two states are then evolved separately.
The first is an eigenstate of the evolution operator (up to some small error given by the variance from vMPS), $\ket{\psi_{t}^{\rm LD}} = e^{t\theta(s)}\ket{\psi_{0}^{\rm LD}}$.
The remaining state is a mixture of all other eigenstates in the spectrum and cannot be easily evolved in the same way.
Fortunately, a range of techniques have been developed for MPS to allow for time-evolution (see Ref.~\cite{Paeckel2019} for comparisons). We will focus on the method introduced in \cite{Verstraete2004b}, with details below.

The time-evolution of a MPS (tMPS) can be achieved by sequentially applying the time evolution operator $U(\delta) = e^{\delta\mathbb{H}_{s}}$ to the MPS. This is approximated using (second order) Trotter-Suzuki decomposition \cite{Suzuki1976} with small times $\delta\ll1$.
We find a Trotter step of $\delta \in [0.01, 0.1]$ to be sufficient.
To avoid exponential growth of the bond dimension, we apply a full truncation scheme through singular value decomposition and variational sweeps to minimize the distance between the truncated and un-truncated MPS, keeping only a maximum of $D = 400$ states (although in practice we never reach this) and with a truncation error $\epsilon = 10^{-12}$ \cite{Schollwoeck2011}.
Note that when we perform the variational truncation, we must also project out the leading eigenvector again, as it may be re-introduced through truncation.
The full time-evolved state is then brought back together by summing the two separate states.
The method is outlined in Fig. 4.

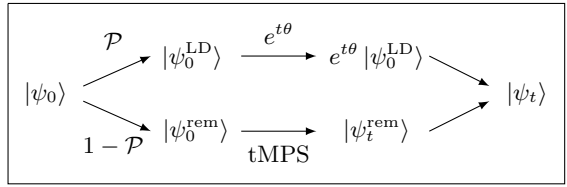
\begin{figure}[h]
\centering
\begin{tikzpicture}
    \draw (-0.5,1.2) -- (6.9, 1.2) -- (6.9, -1.2) -- (-0.5, -1.2) -- (-0.5, 1.2);
    \node (text) at (0, 0) {$\ket{\psi_{0}}$};
    \node (text2) at (1.95, 0.5) {$\ket{\psi^{\rm LD}_{0}}$};
    \node (text3) at (2, -0.5) {$\ket{\psi_{0}^{\rm rem}}$};
    \draw [-latex] (0.5, 0.1) -- (1.4, 0.5);
    \draw [-latex] (0.5, -0.1) -- (1.4, -0.5);
    \node (label) at (0.9, 0.7) {$\mathcal{P}$};
    \node (label) at (0.9, -0.7) {$1 - \mathcal{P}$};
    \node (text4) at (4.4, 0.5) {$e^{t\theta}\ket{\psi^{\rm LD}_{0}}$};
    \node (text5) at (4.4, -0.5) {$\ket{\psi_{t}^{\rm rem}}$};
    \draw [-latex] (2.6, 0.5) -- (3.6, 0.5);
    \draw [-latex] (2.6, -0.5) -- (3.6, -0.5);
    \draw [-latex] (5.1, 0.5) -- (5.9, 0.1);
    \draw [-latex] (5.1, -0.5) -- (5.9, -0.1);
    \node (text5) at (6.4, 0) {$\ket{\psi_{t}}$};
    \node (text7) at (3.1, 0.8) {$e^{t\theta}$};
    \node (text8) at (3.1, -0.8) {tMPS};
\end{tikzpicture}
\label{TEBD-schematic}
\caption{A schematic drawing of the time evolution methods used here. We project the initial state onto the LD vector and the remainder. The former can be evolved exactly as itself up to the exponential pre-factor $e^{t\theta(s)}$, whereas the remainder must be evolved approximately using tMPS. We then add the two back together to give the overall state.}
\end{figure}

We now offer a few comments to the effectiveness of this method.
Firstly, the time evolution is in {\em real time} (by this convention, quantum evolution is imaginary time).
In the large time limit, the leading eigenvector will dominate as the other vectors will be exponentially dampened. The methods here allow us to accurately unravel the contribution of the leading eigenvector up until the times it becomes dominant. 
Furthermore, this approach often allows us to simulate large times by only having small simulation times in tMPS. That is, the remaining state can quickly converge onto the second leading eigenvector (which we determine through the change in norm), allowing us to stop the simulation early and extrapolate to large times.
This is particularly useful when determining the transition from active-to-inactive dynamics for small $s>0$, where timescales diverge but all but the leading two eigenvectors are exponentially dampened. 
Notice that one can adapt this method to allow for multiple leading eigenvectors of the generator, which can be determined variationally. Indeed this could provide a more precise determination of the partition sums, but we find this not to be necessary here.

\subsection{Errors}

Due to the numerical nature of the methods used here, errors are unavoidable.
The first error is introduced due to the approximation of the leading eigenvector. In particular, $\ket{\psi^{\rm LD}}$ has an error which can be measured through the variance with respect to the generator \cite{Banuls2019}. In practice, for the times considered here, this error is small and can be considered negligible.
The dominant sources of error come from our approximation to the remainder, $\ket{\psi^{\rm rem}_{t}}$. This is calculated by evolving the initial state forward in time using a Trotter-decomposed MPO approximation to the evolution operator $U(\delta) = e^{\delta \mathbb{H}_{s}}$.
Here we use a second order Trotter decomposition which entails an error $O\left(N\delta t^{3}\right)$ per time step (and system size $N$), resulting in the accumulated error $O\left(t\delta t ^{2}\right)$ for $M = t / \delta t$ time steps.
Furthermore, after each time step we then truncate the MPS to an upper-bounded bond dimension. In practice, we measure these truncation errors to be very small.
Figure 5 compares the measured (log) partition sum for against numerically exact results for small system sizes (left panel), with the inset showing the error
\begin{equation}
    \delta Z = \left|\frac{\ln(Z^{\rm exact}) - \ln(Z^{\rm MPS})}{\ln(Z^{\rm exact})}\right| .
    \label{error}
    \tag{S6}
\end{equation}
We observe the largest discrepancy around the time where the $\ket{\psi^{\rm rem}_{t}}$ becomes less dominant than $\ket{\psi^{\rm LD}_{t}}$.
At large times of course, the leading eigenvector exponentially dominates and thus the error drops.

\begin{figure}[h]
\includegraphics[width=\linewidth]{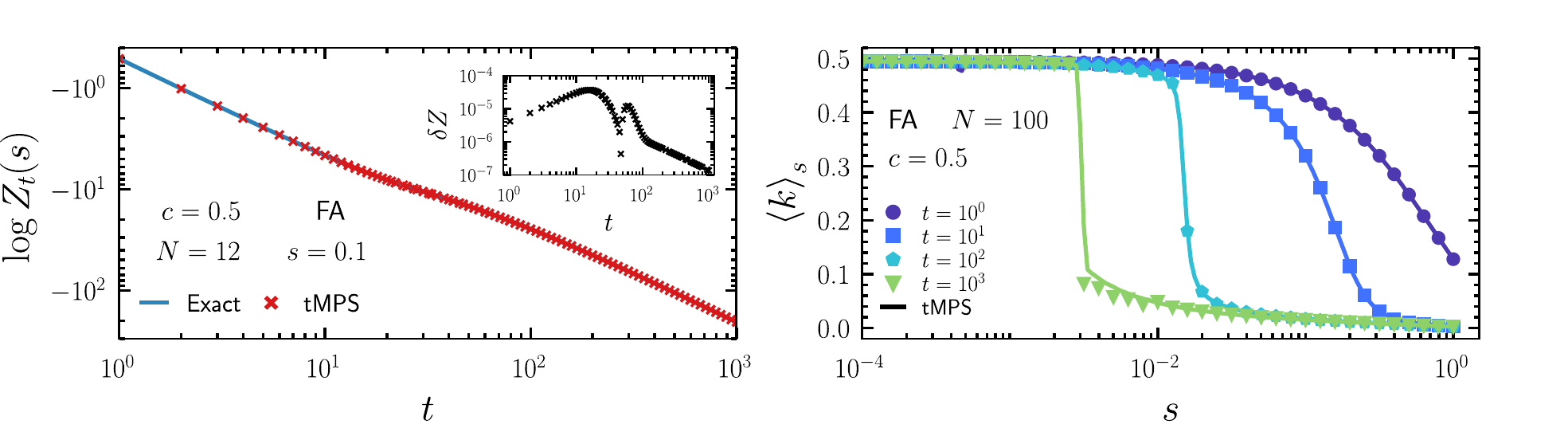}
\centering
\label{fig: errors}
\caption{Demonstration of the errors accumulated using the time evolution scheme here. The left panel compares results from exact numerics (line) and MPS (crosses) for a small system size. The inset shows the error \eqref{error}. The right panel compares the measured activity through determination of the partition sum (line), and sampling using the MPS reference dynamics (symbol) for various times and a large system size $N = 100$.}
\end{figure}

As discussed in the main text, an attempt can be made to correct on some of these errors by using the MPS retrieved after time evolution of \emph{half} the trajectory time as a reference dynamics for umbrella sampling, see Ref.~\cite{Causer2021}. 
Granted enough simulations, if the reference dynamics well approximates the true dynamics, then we could see slight improvements on the measured dynamical activity - if the expected activity from the partition sum largely differs from this result however, it could indicate substantial errors.
The right panel of the figure below shows this comparison for the FA model with $N = 100$. Notice the overwhelming agreement between results, with only small errors around the transition point at large times, although it still correctly predicts the location of the transition point.

\end{document}